\documentclass[twoside]{dis08}
\usepackage[latin1]{inputenc}
\usepackage[dvips]{graphicx,epsfig,color}
\usepackage{wrapfig,rotating}
\usepackage{amssymb,amsmath,array}

\begin{document}
\title{Monte Carlo analysis of CLAS data}

\author{L. Del Debbio$^1$, A. Guffanti$^2$, A. Piccione$^3$
%
%
\vspace{.3cm}\\
%
1- Particle Physics Theory Group, School of Physics, 
University of Edinburgh
\vspace{.1cm}\\
2- Physikalisches Institut, Albert-Ludwigs-Universit\"at, Freiburg
\vspace{.1cm}\\
3- Dipartimento di Fisica, Universit\`a di Milano and 
INFN, Sezione di Milano
\vspace{.1cm}\\
}

\maketitle

\begin{abstract}
We present a fit of the virtual-photon
scattering asymmetry of polarized Deep Inelastic
Scattering which combines a Monte Carlo technique with the use
of a redundant parametrization based on Neural Networks. We apply the 
result to the analysis of CLAS data on a polarized proton target.
\end{abstract}

\section{Introduction}

In High Energy Physics assessing the uncertainty of a function ({\it
  i.~e.}, a probability measure on a space of functions) from a finite
set of experimental data points has become a subject of great interest
in the last years. The case of structure functions in polarized Deep
Inelastic Scattering (DIS) is interesting in this respect, because it
provides a good testing ground for the fitting technique on a
relatively small set of data.

In our analysis we have used the data for the virtual-photon
scattering asymmetry $A_1 (x, Q^2)$, since in this case the theoretical
assumptions are minimized, and then different settings for the
reconstruction of $g_1 (x, Q^2)$ can be tested separately. Indeed,
the extraction of the polarized structure function $g_1 (x, Q^2)$ from
the experimental measured asymmetries $A_{\parallel}$ and $A_{\perp}$
is not so trivial: assumptions have to be made (say, the choice
of a parametrization for the unpolarized structure function) and
different experimental collaborations sometimes quote different
results. Once a fit of $A_1 (x, Q^2)$ is provided, such a
parametrization might be an ideal input for a fit based on
factorization scheme-invariant evolution equations aiming to a precise
determination of $\alpha_s$, as proposed in \cite{BB,BG}.

In \cite{dgp} we have adopted the technique presented in
\cite{Forte:2002fg,Del Debbio:2004qj,DelDebbio:2007ee} to the
available data for polarized DIS. Here we present some details of our
analysis not included in \cite{dgp}, as an example of application
of our results. 

\section{The technique}

The first step is the propagation of experimental information from the
finite set of data points to the parametrization. Since experimental
data are occurrences of a probability distribution, we use a Monte
Carlo technique to sample such a distribution. This sampling is
performed by generating $N_{rep}$ artificial replicas of data
following a multi-gaussian distribution centered on each data point
with the variance given by the experimental error
\begin{eqnarray}
A_1^{(art),k} (x,Q^2)= \left(1+r_{k,N}\sigma_N \right)
\left(A_1^{(exp)}(x,Q^2)+r_{k,t}\sigma_t(x,Q^2)\right)\,,
\end{eqnarray}
where $r$ are Gaussian distributed random numbers, $\sigma_t$ is is
the total uncorrelated error, obtained by summing in quadrature the
statistical and uncorrelated systematic errors, $\sigma_N$ is the
normalization multiplicative error. At this stage of the analysis we
neglect correlated systematic errors since they are not available for
all the experiments considered. Moreover the measurement of polarized
structure functions is dominated by statistical errors.

The number of Monte Carlo replicas of the data is determined by the
requirement that the average over the replicas reproduces the features 
(central values and errors) of the original experimental 
data with a given accuracy. In practice here we use $N_{rep}=100$.

The second step is to build a representation of the probability
measure in the space of asymmetry functions. This construction can be
done by taking a given functional form, and determining its parameters
by fitting each of the generated replicas. In order to minimize the
bias due to the assumptions made on the shape of the fitted functions
we make use of a redundant parametrization, that is a parametrization
with a sufficiently large number of parameters so that not only the
underlying physical law, but also the statistical fluctuations of the
data could be reproduced. Of course within this framework over-fitting
is allowed and the minimization becomes crucial. In practise we use a
Neural Network which has 20 tunable parameters to perform a fit to 100
data points and we stop the minimization, when we start fitting
statistical noise (see \cite{dgp} for details).

At the end of this procedure any observable can be computed from the fit
by averaging over the sample. As an example, if we neglect the contribution
of $g_2 (x, Q^2)$, we have
\begin{eqnarray}
g_1 (x, Q^2) \simeq \frac{1}{N_{rep}}\sum_{k=1}^{N_{rep}} A_1^{(fit),k}(x, Q^2) F_1 (x, Q^2)\,.
\end{eqnarray}

\section{Results}

\begin{center}
\begin{figure}
\centerline{\includegraphics[width=0.65\columnwidth]{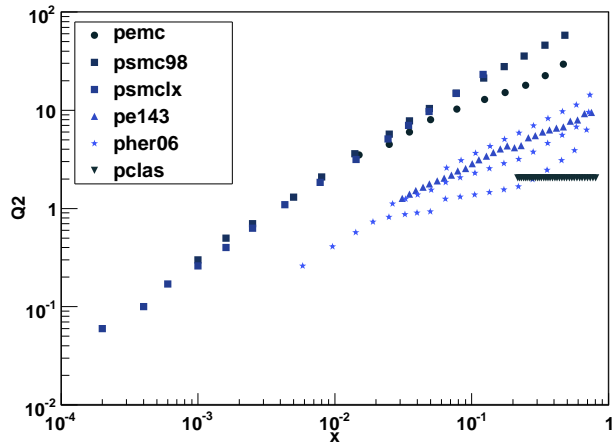}}
\caption{Kinematics of the experimental data used in the analysis plus
the CLAS bin used to test our fit.}
\label{Fig:Kinematics}
\end{figure}
\end{center}

As an application and a test of our result, we compare our extraction
of $g_1 (x, Q^2)$ to CLAS data \cite{Dharmawardane:2006zd}, which have
not been used in the fit. In Figure \ref{Fig:Kinematics} we show the
kinematic range spanned by the data used in the fit (see also Table
\ref{tab:data}) together with the bin of CLAS data against which we
want to check our parametrisation. As it can be easily seen most of
the CLAS data are in a region where our fit is asked to extrapolate,
both in $x$ and in $Q^2$.

\begin{center}
\begin{figure}
\includegraphics[width=0.49\columnwidth]{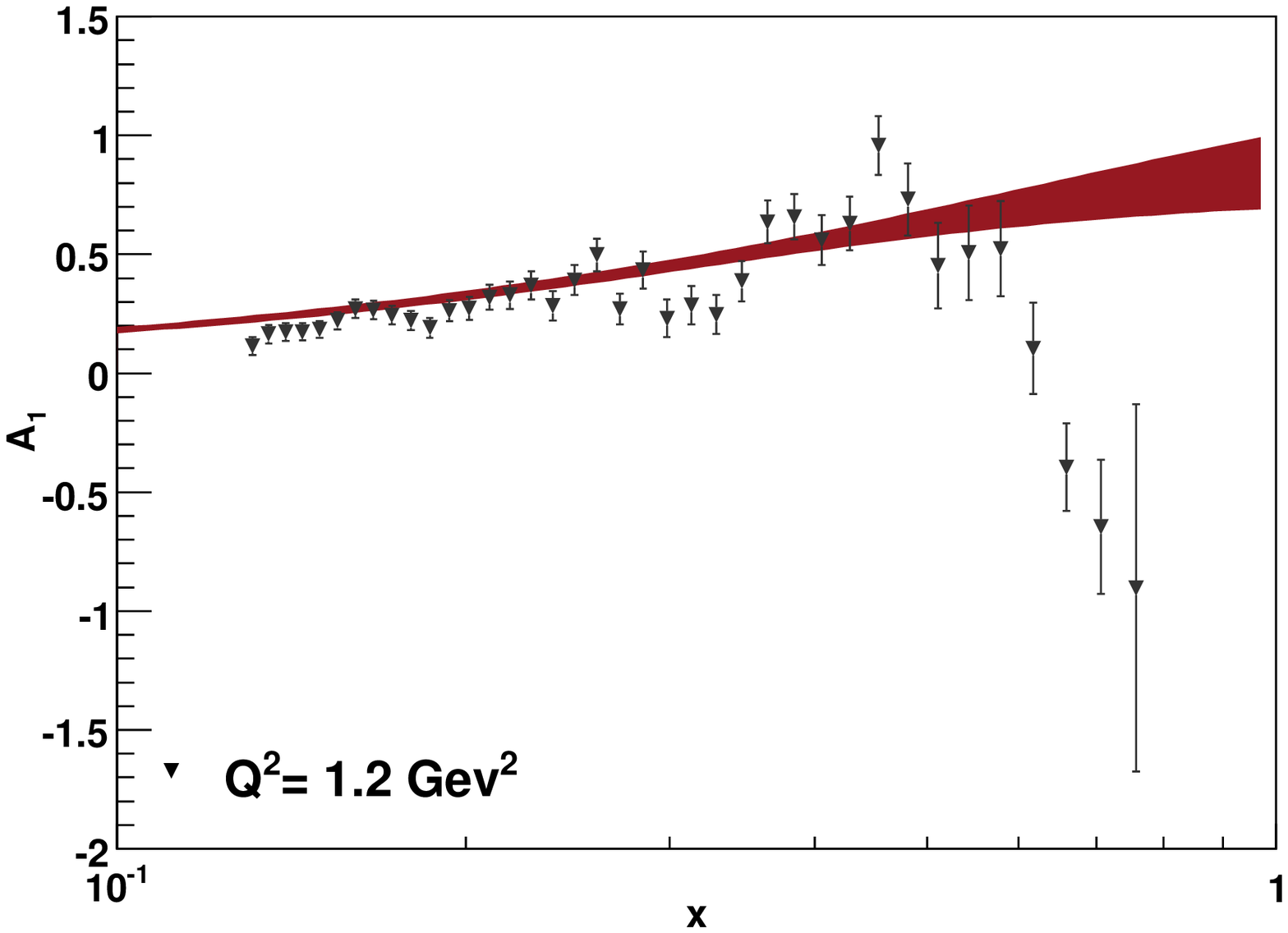}
\includegraphics[width=0.49\columnwidth]{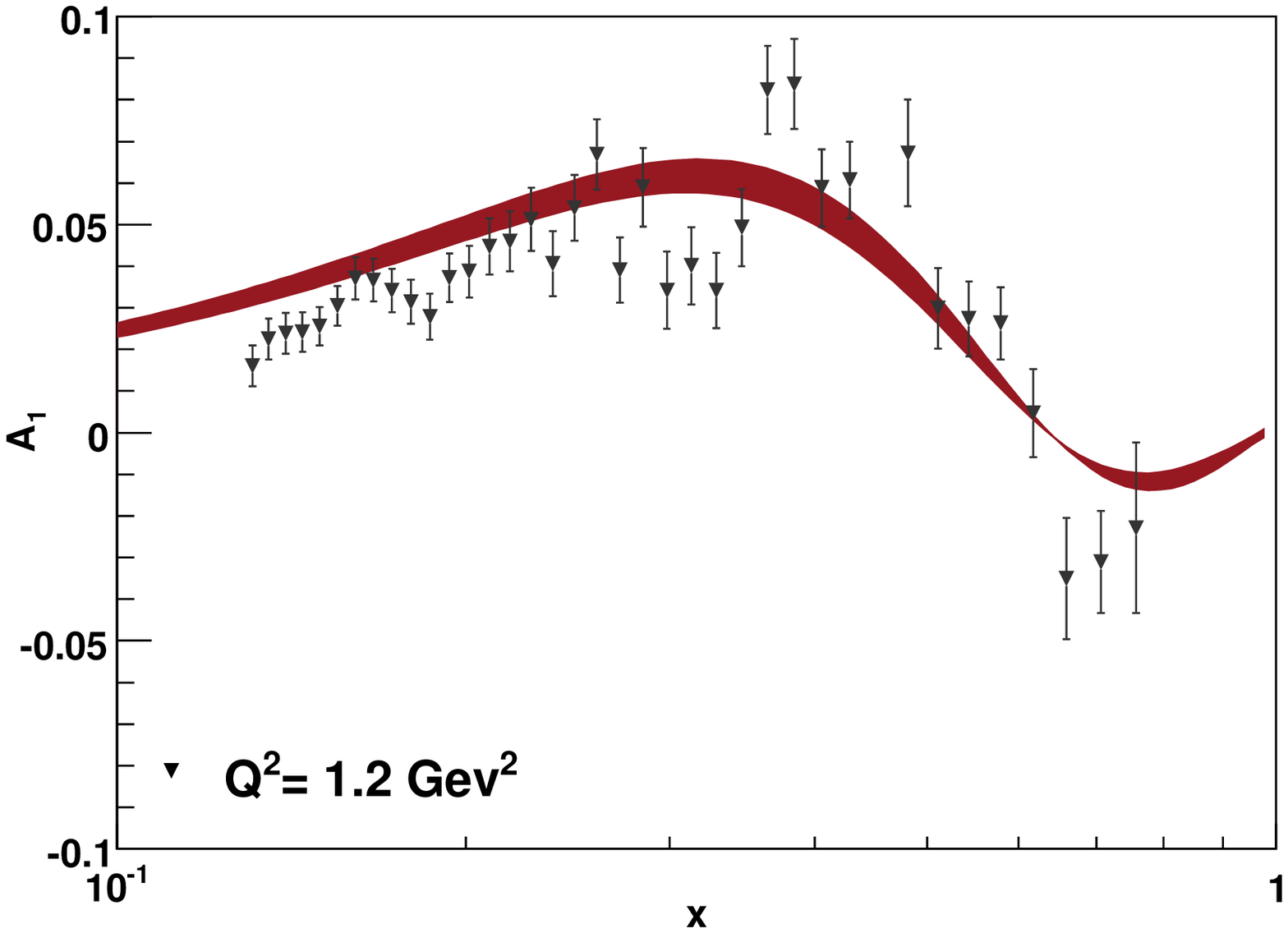}
\caption{The fitted asymmetry $A_1 (x, Q^2)$ and  reconstruction of $g_1(x,Q^2)$ 
compared to CLAS data not used in the fit}
\label{Fig:g1}
\end{figure}
\end{center}

Comparing our extraction of $A_1 (x, Q^2)$ with data (see Figure
\ref{Fig:g1} left pad), we notice that our description and CLAS data are
compatible within 2-$\sigma$ error band, and that resonances are
naturally averaged by the fit.

\begin{table}
\begin{center}
\begin{tabular}{|l|ll|c|} 
\hline
Proton & $x$ range & $Q^2 (GeV^2)$ range & $N_{dat}$ \\
\hline
\hline
EMC & 0.0150 - 0.466 & 3.50 - 29.5 & 10 \\
\hline
SMC & 0.0010 - 0.480 & 0.30 - 58.0 & 15 \\
\hline
SMC low-$x$ & 0.0001 - 0.121 & 0.02 - 23.1 & 15 \\
\hline
E143 & 0.0310 - 0.75 & 1.27 - 9.52 & 28 \\
\hline
HERMES06 & 0.0058 - 0.7311 & 0.26 - 14.29 & 45 \\
\hline
\end{tabular}
\caption{General features of experimental data used in the fit}
\label{tab:data}
\end{center}
\end{table}

In order to obtain the polarized structure function $g_1 (x, Q^2)$, we
have to supplement our parametrization with some assumptions, whose
impact can be tested easily once a parametrization of $A_1 (x, Q^2)$
is given. In the present analysis we take $g_2 (x, Q^2) = g_2^{WW}
(x, Q^2)$, and
\begin{eqnarray}
F_1(x, Q^2) = \frac{1+\gamma^2}{2x\,(1+R(x, Q^2))}F_2(x, Q^2)\,
\end{eqnarray}
where $\gamma^2=4xM^2/Q^2$, $R(x, Q^2)$ is given from the SLAC
parametrization \cite{Whitlow:1990gk,Abe:1998ym}, and $F_2(x, Q^2)$
from \cite{Del Debbio:2004qj}. We have
\begin{eqnarray}
g_1 (x, Q^2)=  \frac{1+\gamma^2}{2x\,(1+R(x, Q^2))} 
\frac{1}{N_{rep}}\sum_{k=1}^{N_{rep}} \left [A_1^{(fit),k}(x, Q^2) F_2^{k} (x, Q^2)
+ \gamma^2 g_2^{k}(x, Q^2)\right]\,,
\label{eq:g1}
\end{eqnarray}
with
\begin{eqnarray}
g_2^{k}(x, Q^2)=-g_1^{k}(x, Q^2)+ \int_x^1 \frac{dy}{y}\,g_1^{k}(x, Q^2)\,.
\label{eq:g2WW}
\end{eqnarray}
Equations (\ref{eq:g1}) and (\ref{eq:g2WW}) explicitly show how to
evaluate any quantity depending on the asymmetry, together with its
associated error, in a straightforward way. The result shown
in the right pad of Figure \ref{Fig:g1} is consistent with the one
given in \cite{Bosted:2006gp}.

We have presented a fit based on a large fraction of the available
experimental data for the asymmetry $A_1 (x, Q^2)$, and we have
applied this result to predict some of the remaining data that were
not included in the fit and lie in a region where our asymmetry needs
to be extrapolated. We find the result of this comparison to be
satisfactory. We conclude that our parametrization provides an
interesting tool to study quark-hadron duality since no QCD evolution
is used and averages of the structure function over resonances are
naturally guaranteed by the smoothness of the fit (see
\cite{Bosted:2006gp} and references therein for further details on a
usual approach to quark-hadron duality).

\section{Acknowledgments}

LDD is supported by an STFC advanced fellowship.


\begin{footnotesize}




\begin{thebibliography}{99}
\bibitem{url} Slides: \\ 
\verb$http://indico.cern.ch/materialDisplay.py?contribId=266&sessionId=22&materialId=slides&confId=24657$

\bibitem{BB} J.~Bl\"umlein and H.~B\"ottcher,
	     Nucl.\ Phys.\ B {\bf 636} (2002) 225.

\bibitem{BG} J.~Bl\"umlein and A.~Guffanti,
	      AIP Conf.\ Proc.\ {\bf 792} (2005) 261.

\bibitem{dgp} L.~Del~Debbio, A.~Guffanti, A.~Piccione, in preparation.

\bibitem{Forte:2002fg}
  S.~Forte, L.~Garrido, J.~I.~Latorre and A.~Piccione,
  JHEP {\bf 0205} (2002) 062
  [arXiv:hep-ph/0204232].

\bibitem{Del Debbio:2004qj}
  L.~Del Debbio, S.~Forte, J.~I.~Latorre, A.~Piccione and J.~Rojo  [NNPDF
                  Collaboration],
  JHEP {\bf 0503} (2005) 080
  [arXiv:hep-ph/0501067].

\bibitem{DelDebbio:2007ee}
  L.~Del Debbio, S.~Forte, J.~I.~Latorre, A.~Piccione and J.~Rojo  [NNPDF
                  Collaboration],
  JHEP {\bf 0703} (2007) 039
  [arXiv:hep-ph/0701127].

\bibitem{Dharmawardane:2006zd}
  K.~V.~Dharmawardane {\it et al.}  [CLAS Collaboration],
  Phys.\ Lett.\  B {\bf 641} (2006) 11
  [arXiv:nucl-ex/0605028].

\bibitem{Whitlow:1990gk}
  L.~W.~Whitlow, S.~Rock, A.~Bodek, E.~M.~Riordan and S.~Dasu,
  Phys.\ Lett.\ B {\bf 250} (1990) 193.

\bibitem{Abe:1998ym}
  K.~Abe {\it et al.}  [E143 Collaboration],
  Phys.\ Lett.\ B {\bf 452}, 194 (1999)
  [arXiv:hep-ex/9808028].

\bibitem{Bosted:2006gp}
  P.~E.~Bosted {\it et al.}  [CLAS Collaboration],
  Phys.\ Rev.\  C {\bf 75} (2007) 035203
  [arXiv:hep-ph/0607283].
\end{thebibliography}
%

\end{footnotesize}


\end{document}